\PassOptionsToPackage{cmyk,table}{xcolor}
\documentclass[conference,flushend]{iaria} % Optional: flushend or pbalance
\pdfoutput=1 % pdflatex hint for arxiv.org (within first 5 lines)

\usepackage[babel=true,english=american]{csquotes}
\usepackage[USenglish]{babel}

\usepackage{array}
\usepackage{enumerate}
\usepackage{amsmath,amsthm,amsfonts}
\usepackage{caption}
\usepackage{biblatex}
\usepackage[%
	babel=true, % Enable language-specific kerning. Take language-settings from the languge of the current document (see Section 6 of microtype.pdf)
	expansion=alltext,
	protrusion=alltext-nott, % Ensure that at listings, there is no change at the margin of the listing
	nopatch=eqnum, % fix unable to apply patch eqnum
	final % Always enable microtype, even if in draft mode. This helps finding bad boxes quickly.
]{microtype}

\newcolumntype{C}[1]{>{\centering\arraybackslash}p{#1}}

\title{An Approach for Decentralized Authentication in Networks of UAVs}

\author{\large Nicholas Jäger and Andreas Aßmuth\,\orcidlink{0009-0002-2081-2455}\\[0.3ex]\normalsize\normalfont
	Technical University of Applied Sciences OTH Amberg-Weiden\\
	Department of Electrical Engineering, Media and Computer Science\\
	Kaiser-Wilhelm-Ring 23, 92224 Amberg, Germany\\
	{\tt \{n.jaeger$\,|\,$a.assmuth\}@oth-aw.de}%
}

\usepackage{ifpdf}
\ifpdf
\pdfoutput=1 % we are running pdflatex
\pdftrue
\fi

\makeatletter
\def\ps@IEEEtitlepagestyle{
  \def\@oddfoot{\mycopyrightnotice}
  \def\@evenfoot{}
}
\def\mycopyrightnotice{
  {\footnotesize
    \begin{minipage}{0.8\textwidth}
    \centering
    Please cite as: \fullcite{selfref}.
    \end{minipage}
  }
}
\makeatother
\makeatletter
\let\blx@rerun@biber\relax
\makeatother
\usepackage[shortcuts]{extdash} % Use \-/ for a breakable dash that does not prevent the remainer of the word to be hyphenated

\DeclareBibliographyCategory{selfref}
\addtocategory{selfref}{selfref}
\begin{document}

\maketitle

\begin{abstract}
We propose a decentralized authentication system for networks of unmanned aerial vehicles. A blockchain-based public key infrastructure allows the usage of public key cryptography and public key based authentication protocols. The blockchain provides a common storage of the public keys and their relations and can provide the required information for the authentication process. Furthermore, the unmanned aerial vehicles store selected parts of the blockchain in order to operate independently in areas where they might not have access to the Internet. This allows unmanned aerial vehicles to authenticate entities of the network, like other unmanned aerial vehicles, cloud services, cars, and any computer.
\end{abstract}

% A list of IEEE Computer Society appoved keywords can be obtained at
% http://www.computer.org/mc/keywords/keywords.htm

\section{Introduction}
Unmanned Aerial Vehicles (UAVs) have become popular recently in the civilian area because of technological advancement and their great potential for different applications. UAVs can perform a big variety of missions either controlled remotely or in an autonomous fashion. Some of the applications are, for example, delivery of goods, search and rescue missions, wildlife and terrain monitoring, providing emergency infrastructures and many more (see, e.g., \cite{shakhatreh_unmanned_2019}, \cite{bekmezci_security_2016}).\par
The potential of the UAVs is further increased when they are forming networks to share information or to cooperate on a common mission. Due to the open nature of these networks in the civilian domain they are vulnerable to different attacks \cite{hartmann_uav_2016} and must therefore be secured properly.\par
In these networks, the UAVs interact with other UAVs, different kinds of vehicles, infrastructural elements, or diverse cloud services. For security in these networks, the protection goals of authenticity and integrity, among others, must be ensured. This must be ensured when UAVs provide sensor data for further processing in the cloud, for example, in the context of search and rescue missions, wildlife monitoring or collection of current weather data. The same is true in the case when cloud services supply the UAVs with data like maps, no-fly zones, proposed trajectory, or command and control data. It is, therefore, necessary to prevent UAVs and cloud services from compromising each other.\par
Secure communication starts with secure authentication. One important security measure is the Public Key Infrastructure (PKI), which allows the secure usage of public key cryptography. It provides the possibility to authenticate entities in a trustworthy way since it binds public keys to entities. A common approach for PKIs are hierarchical PKIs where trusted third parties guarantee the bond between public keys and entities. The trusted third parties can, however, be a single point of failure and can, hence, be considered as a weakness. Decentralized approaches like peer-to-peer PKIs are alternatives to the hierarchical PKIs. The blockchain technology has added new possibilities to design decentralized PKIs, which is why they are attractive alternatives to hierarchical PKIs.\par
In this paper, we propose a blockchain-based PKI for UAVs. The remainder of the paper is organized as follows: in Section~\ref{sec:related-work}, a selection of related work is presented. Subsequently, an overview of the relevant technology is given in Section\ref{sec:background}. Further in Section~\ref{sec:proposed-approach}, the design of the blockchain-based PKI is proposed. The research project ADACORSA is briefly introduced in Section~\ref{sec:adacorsa}. Section~\ref{sec:concl} concludes the paper and states what is planned in the future.

\section{Related Work}\label{sec:related-work}
Recently, different solutions for the authentication in networks of UAVs have been proposed. For example, Rodrigues et. al~\cite{rodrigues2019} adapted authentication protocols from the area of wireless sensor networks for the use in networks of UAVs. They use the ground control stations as trusted third parties where the UAVs are registered. Thompson and Thulasiramen~\cite{thompson2016} proposed to use symmetric key cryptography for the communication in swarms of UAVs because of the better performance. The symmetric key has to be preloaded to the UAVs before the mission and the swarm of UAVs forms a closed network.\par
Blockchain technology has already been used to design blockchain-based PKIs and authentication systems in different domains, including networks of UAVs. For example, Yakubov et al. use blockchain technology to improve existing PKIs systems like the peer-to-peer PKI used in PGP~\cite{yakubov_2018b} and hierarchical PKIs based on X.509 certificates~\cite{yakubov2018a}. An overview of blockchain-based PKIs can be found in~\cite{salman2019}.\par
For example, Yazdinejad et al.~\cite{yazdinejad_enabling_2020} utilized blockchain technology to develop an authentication system for UAVs in smart cities which are divided into zones. For every zone, a zone controller is responsible and logs its activities on a public blockchain. The UAVs have to register at a zone controller. It assigns cryptographic keys to the UAV and logs the data of the drone in the blockchain.\par
In this paper, we use the blockchain technology to design a decentralized PKI, i.d., trusted third parties are not required, for open networks of UAVs and the Internet of UAVs.

\section{Background}\label{sec:background}
In this section a short overview of the different relevant technologies is given: In Subsection~\ref{subsec:network}, the characteristcis of networks of UAVs are described. Blockchain technology is presented in Subsection~\ref{subsec:blockchain} and PKIs and their trust models in Subsection~\ref{subsec:pki}.

\subsection{Network of UAVs}\label{subsec:network}
Wireless communication technology enables the UAV to communicate with different entities: with the ground station their operator, with other UAVs, with other types of vehicles, and, possibly, other services in a private or public cloud. By forming Flying Ad-hoc Networks (FANETs), UAVs can exchange information and cooperate in order to fulfill their mission. If the UAVs are connected to the Internet, their network is expanded to the Internet of UAVs as a part of the Internet of Things.\par
FANETs are a subset of Mobile Ad-hoc Networks (MANETs) and share some of their characteristics but also differ in some aspects. FANETs are characterized by a high mobility of their nodes, a continuously changing topology, a low node density, and limited available resources of the nodes like power, memory, and computational power. Therefore, security solutions of the MANET domain cannot be adopted without risking that they become less efficient or even fail \cite{bekmezci_security_2016}.

\subsection{Blockchain}\label{subsec:blockchain}
Blockchain technology, introduced by the bitcoin protocol~\cite{bitcoin2008} in 2008, allows agreement on a common state of a system in an open network in a decentralized manner, i.e., without using trusted third parties or intermediaries. The term blockchain has two different but related meanings. In the first meaning, it denotes a special data structure whose elements, the blocks, are connected by cryptographic hash functions. A block consists of a block header with meta data and a list of transactions,i.e., the content. The list of transactions is cryptographically linked to the block header, e.g., by a Merkle tree~\cite{merkle1988}. In the second meaning, it describes a system in which this data structure is distributed in a (peer-to-peer) network and an associated protocol that prescribes how new data can be added and agreed upon (consensus process). The protocol allows only to append new data and it should be impossible to delete blocks that the network has agreed on. For an overview of consensus protocols we refer exemplary to~\cite{nguyen2018}.\par
One can distinguish between different kinds of blockchain systems~\cite{zheng2017}: In a public blockchain, everyone can read the stored data and can participate in the consensus process, in principle. In a consortium blockchain, a selected group is allowed to attend the consensus process. The stored data may be read by selected members or by the public. In a private blockchain, all participants belong to the same organization and the system cannot be accessed by the public.\par
The advantage of blockchain systems from a security point of view is that they can guarantee the integrity and availability of the stored data~\cite{salman2019}. Since blockchain systems can be very transparent and their state can be observed and checked by the participants, they do not require much trust in each other. Therefore, it is possible to store data generated by the blockchain, e.g. blockchain tokens, in a trustworthy manner. However, additional measures must be taken to ensure that other kinds of data that does not stem from the blockchain can be trusted.

\subsection{Public Key Infrastructures and Trust models}\label{subsec:pki}
PKIs allow the secure usage of public key cryptography by binding public keys to identities in a trustworthy manner~\cite{buchmann2013}. Usually PKIs issue certificates which confirm that the mentioned identity controls the associated keys. Furthermore, they also manage, distribute, and sometimes revoke certificates. The trust model (or authentication metric) of the PKI defines the set of rules to accept certificates. Two classes of PKIs can be distinguish: hierarchical PKIs and peer-to-peer PKIs.\par
Hierarchical PKIs are categorized by the fact that only special entities, the so called Certificate Authorities (CAs) have the right to issue and revoke certificates to other entities, including other CAs. Since these CAs can also issue certificates, a hierarchy of CAs emerges. A CA which is not certified by another CA is called Root-CA and serves as a trust anchor; the other CAs are called intermediary CAs. Hence, the certified entities are connected by a chain of certificates to the Root-CA. The security of certificates (and the chain of certificates) depends on the trustworthiness of the issuing CAs and the users have to rely on CAs to carefully verify the claimed relationships between the identities and keys.\par
The relationships in a PKI can be mathematically modeled as a directed graph, called trust graph, where the nodes represent the entities and their public keys and the edges represent certificates between the entities. In a hierarchical PKI, the graph has the form of a tree~\cite{buchmann2013}. The leaves of the tree represent the end-entities and the root of the tree represents the Root-CA and the intermediary nodes represent the intermediary CAs.\par
In a simple trust model the Root-CA acts as the trust anchor of the tree, i.e., all nodes directly trust the root and, hence, indirectly trust all other nodes. Even entities which are not part of the tree can decide to trust the root and, therefore, also any node of the tree. A hierarchical PKI is not limited to a single Root-CA (and a single tree), but can have several independent Root-CAs and, hence, several trust anchors. There are several methods to connect the different trees to each other, e.g., it might be sufficient that the user decides to directly trust a set of the different trust anchors. Another possibility is to introduce a new Root-CA and subordinate the existing ones. Cross certification (roots certify each other) or bridges (a new node that is cross-certified by several Root-CAs) are options without subordination~\cite{perlman1999}. In this trust model, the process of validating a key-identity-binding consists in finding a path from the entity to a trust anchor, i.e., finding a certificate chain.\par
In peer-to-peer PKIs, everyone has the right to issue and revoke certificates and, hence, users may directly trust each other. The graph of a peer-to-peer PKI is usually not a tree but has a more complex structure and might be better described by other network models like small world graphs or scale-free networks. The validation of a key-identity-binding requires finding a trustworthy path from the own node to the node of the communication partner. The associated trust model must contain a mechanism to evaluate the trustworthiness of a path.\par
For an overview of the different kinds of hierarchical and peer-to-peer trust models we refer exemplary to \cite{buchmann2013, perlman1999, maurer1996, reiter1999, alcalde2010}.

\section{Proposed Approach for a PKI for network of UAVs}\label{sec:proposed-approach}
In this section, we propose a blockchain-based public key infrastructure for the networks of UAVs for the prupose of authentication. In Subsection~\ref{subsec:overview} the basic idea of our proposal is introduced. The different components of the system are described in the following subsections: the design of the blockchain in Subsection~\ref{subsec:bc-design}, the proposed trust model of the PKI in Subsection~\ref{subsec:trust-model} and the distribution of the data in Subsection~\ref{subsec:auth-data}. Finally, the authentication process is outlined in Subsection~\ref{subsec:auth-proc}.

\subsection{Overview}\label{subsec:overview}
\begin{figure}[!ht]
	\centering
	\includegraphics[width=8cm]{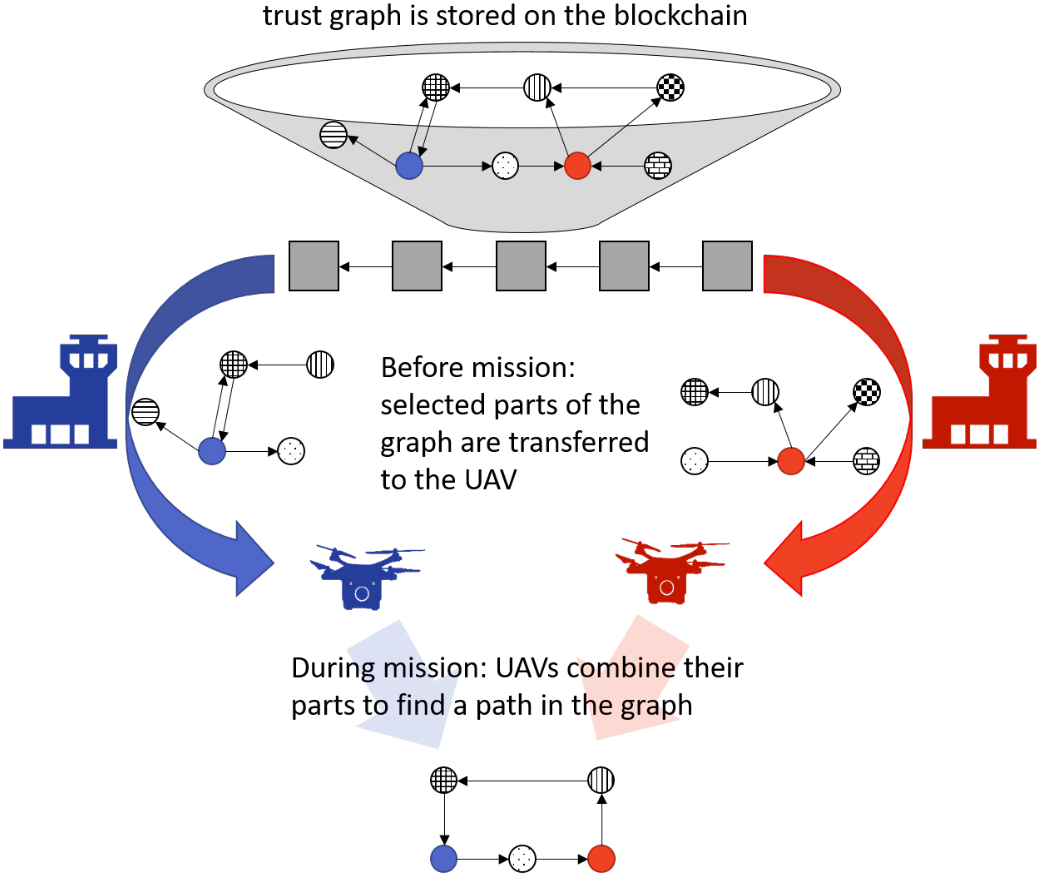}
	\caption{Schematic overview of the approach.}
	\label{fig:overview}
\end{figure}

The basic concept of this approach is to store the public keys, the identities, and their trust relationships in a dedicated public blockchain. Therefore, the blockchain contains the trust graph of the PKI. For this purpose, the blockchain offers special transactions. Due to their limited resources, the UAVs do not participate as nodes in the blockchain system and do not store the whole blockchain. They store only the part of the blockchain which is relevant to them. During the authentication process the two UAVs combine their knowledge to find a trustworthy path in the trust graph. This idea is depicted in Figure~\ref{fig:overview}.

\subsection{Blockchain Design}\label{subsec:bc-design}
We propose a dedicated public blockchain system, i.e., everyone is allowed to join the blockchain network to participate in the consensus process, with an appropriate consensus protocol because a dedicated blockchain system can be designed to fit the needs of a PKI. Here, we do not specify the blockchain design in detail, but we describe some elements of the system from a high-level point of view. For the sake of clarity, we
assume a blockchain design that is based on the Bitcoin blockchain~\cite{bitcoin2008}. Therefore, its consensus protocol (e.g., Proof of Work, Proof of Stake, etc.) uses tokens as a currency to reward the nodes which participate in the process.\par
In a blockchain system, transactions are used to change the state of the system, e.g., adding new information, and the allowed set of transactions defines the capabilities of the system. Since we propose to use a dedicated blockchain, we can choose a set of transactions which provides the required functionality. Therefore, the flexibility provided by transactions that allow to store and execute programs, so called smart contracts, are not needed and, hence, it is not necessary that the blockchain system offers such transactions. In contrast to the Bitcoin blockchain~\cite{bitcoin2008} we assume that the transactions are account-centered and do not require to reference to previous transactions and do not offer or need a scripting mechanism to release it. A transaction, therefore, usually consists of the following elements:
\begin{itemize}
	\item A sender, i.e., the account issuing the transaction.
	\item A workload which contains information like the receiver of the transaction.
	\item Formal elements, like the type, id, hash of the transaction and cryptographic mechanisms that verify that the sender has authorized the transaction.
\end{itemize}
Since the blockchain uses tokens as currency, a transaction to create tokens, the so called coinbase transaction, and a transaction to transfer tokens are needed. Furthermore, a transaction is required to store data that represent an entity, containing its name, public key, and maybe also some of its characteristic properties like its type, model, and the responsible authority. This type of transaction creates a node of the trust graph. Transactions that confirm bindings between keys and entities fulfill the task of certificates and correspond to the creation of edges in the trust graph. The deletion of edges, i.e., the revocation of certificates, is performed by transactions that nullify previous given confirmations. These three types of transactions are sufficient to store the trust graph of a peer-to-peer PKI on the blockchain. Additional transactions can extend or optimize the functionality but are not considered here.\par
Furthermore, it is desirable that the blockchain system provides a secure mechanism to create checkpoints of the blockchain state. By checkpoint, we mean a data structure that stores the state of the blockchain at a given block. A checkpoint $c_m$ at block $B_m$ which is at the position $m$ in the chain should have the property that the checkpoint together with the blocks $B_{m+1}, B_{m+2}, ... B_{m+n}$ is sufficient to obtain the state of the blockchain at block $B_{m+n}$. The checkpoint does not need to store the history of the system but only the results, e.g., it only stores the balance of an account and not changes of the balance. Therefore, a checkpoint can be used to get a compressed version of the blockchain.
A checkpoint could be realized by a transaction that contains a reference to the block $B_m$ and the associated checkpoint $c_m$. Together with a protocol defining the creation of a checkpoint, the nodes can verify the correctness of the checkpoint and, hence, it can be used in future.

\subsection{Trust model}\label{subsec:trust-model}
For this PKI we propose to use the following peer-to-peer trust model which is based on~\cite{maurer1996}: Everyone is allowed to create a transaction which binds its identity to its public keys. They can also confirm the binding between identity and public keys of other user and revoke their previous given confirmation. When a user $A$ confirms another entity $B$ they assign a number $n \in \{1, 2, ..., m\}$ to this relation where $m\in\mathbb{N}$ denotes a global parameter of the trust model, we write:
\begin{align*}
	A \overset{n}{\rightarrow} B.
\end{align*}
This number means the maximal length of the path starting with the edge $(A,B)$ which the user is willing to accept: $n=1$ means that $A$ only trusts $B$; $n=2$ that it might also trust all entities which are confirmed by $B$ and so on. Furthermore, the path has to respect all numbers of the path, i.e., a partial path can only be as long as the number of its starting edge is allowing. For example, we evaluate the situation
\begin{align*}
	A \overset{3}{\rightarrow} B \overset{1}{\rightarrow} C \overset{2}{\rightarrow} D.
\end{align*}
Even though $A$ accepts paths of length $3$ starting with the edge $(A,B)$, $B$ only allows a path of length $1$ starting with the edge $(B,C)$.
Therefore, the path $A-B-C-D$ is not allowed.\par
We have chosen this trust model since it incorporates the facts that trust is not transitive in general,
\begin{align*}
	A \rightarrow B, B \rightarrow C \not\Rightarrow A \rightarrow C,
\end{align*}
and it is reducing with growing distance. Furthermore, it is simple and does not require the evaluation of parallel paths (e.g., $A-B_1-C$ and $A-B_2-C$) in order to determine the trustworthiness of an identity-key-binding.

\subsection{The data for authentication stored by the UAVs}\label{subsec:auth-data}
The UAVs only have limited capabilities to store and process data. Therefore, the UAVs can neither participate in the blockchain network nor store the whole blockchain. They only require the nodes and edges of the trust graph they are trusting and only have to store a selection of the blockchain data, e.g., the headers of the blocks and the transactions which are relevant for their view of the trust graph. The relevant part of the trust graph may still be too big for the UAV, but it can still be reduced by the fact that every UAV has to store a part of the trust graph and can exchange their parts in the authentication process. Assuming that a node has $n$ trusted neighbors on average, it has to store about $n^m$ nodes and edges to reach all trusted nodes within the distance of $m$. But if every node stores all nodes and edges of incoming and outgoing paths of length $k$, which would be $2n^k$ nodes and edges, and combine its stored part with the communication partner, they can reconstruct paths of the length $2k$.\par
The trust graph can further be reduced by considering the trust model and by utilizing the global view on the trust graph, provided by the blockchain. Furthermore, we expect that the UAVs do not primarily confirm other UAVs, but confirm nodes representing the organization which controls the UAVs. Additionally, organizational nodes will confirm other organizational nodes, cloud services, and, therefore, the trust graph will have many hubs.\par
Even though there are already algorithms for distributing trust graphs (see, e.g., \cite{hubaux_quest_2001}, \cite{capkun_self-organized_2003}), we are still working on the development of an algorithm utilizing these aspects.\par
We assume that the operators or ground stations provide their UAVs with the required data before the mission and, hence, the UAV do not have to process the blockchain by themselves. Because of the limited operation time the UAVs should have a rather recent view on the trust graph during their mission.\par
Alternatively, the UAVs can further reduce the amount of data if it can be ensured that the UAV has access to the Internet during the whole mission. In this case they could request the required data from the blockchain network.

\subsection{Authentication process}\label{subsec:auth-proc}
Well-known public key authentication protocols can be adapted for the authentication process. We refer to~\cite{boyd_protocols_2003} as an overview. Here we sketch this process from a high-level point of view: Alice and Bob are two entities (UAVs) and Alice wants to authenticate Bob.
\begin{enumerate}
	\item Bob sends Alice a message with his identity and with a list of hashes of the nodes of his incoming paths.
	\item Alice compares this list with the hashes of nodes of her outgoing paths. When she finds a common hash, she requests the data of the nodes and edges from Bob. In case she does not find a common hash, the authentication process is aborted.
	\item Bob sends the requested data and Alice checks the integrity of the received data with their blockchain headers and their Merkle trees. Then, she reconstructs the path and verifies that it is valid. If one of the checks is negative, the process terminates.
	\item Alice can now use the public key of Bob to authenticate Bob as prescribed in the used authentication protocol.
\end{enumerate}

\section{The research project ADACORSA}\label{sec:adacorsa}
The goal of the project Airborne Data Collection on Resilient System Architectures (ADACORSA)~\cite{adacorsa} is to develop the technical components (hardware, software, etc.) to enable civilian UAVs to operate semi-autonomously beyond the visual line of sight. The project does not deal with UAVs for the military domain. To achieve this goal, work in different domains will be carried out. For example, the required electronics
components for the safe and reliable flight beyond the visual line of sight will be developed, measure to increase social acceptance of civilian UAVs will be conducted. Furthermore, solutions will be designed to secure the communication of UAVs with different parties, like other UAVs, the ground stations, the operators, and other entities, especially in the area of identification and authentication. The project started in May 2020 and will last till May 2023 and brings 49 companies from different domains, research institutes and universities from 12 countries together.

\section{Conclusion and Future Work}\label{sec:concl}
In this paper, we have presented an approach to design a blockchain-based peer-to-peer PKI for UAVs. The blockchain serves as a secure decentralized storage for the trust graph of the PKI and grants a global view. The UAVs do not store the whole blockchain, but only parts of it and combine their knowledge of the trust graph to find a path between them.\par
However, here we have only specified the core concepts of such a PKI, and several steps still have to be taken: An algorithm which selects the relevant parts of the trust graph has to be developed and evaluated in an apropriate context. For this purpose, a method must be developed to generate random trust graphs. The performance of selection algorithm is then analyzed by applying it to random trust graphs of different size and structure. Subsequently, a proof of concept system must be implemented. A proof of concept system could consist of a network of single board computers, like Raspberry Pis, representing the UAVs, and more powerful computers representing ground stations and cloud services. Generally, we
propose using a simple trust model which could be substituted by other ones and their performance can be compared in order to find the most appropriate one.

\section*{Acknowledgments}
This work is supported by ECSEL Joint Undertaking (JU) through the Project ADACORSA under grant agreement No~876019. The JU receives support from the European Union’s Horizon~2020 research and innovation programme and Germany, Netherlands, Austria, Romania, France, Sweden, Cyprus, Greece, Lithuania, Portugal, Italy, Finland, Turkey.

\renewcommand*{\bibfont}{\footnotesize}
\setlength{\labelnumberwidth}{0.45cm}
\printbibliography[notcategory=selfref]

\end{document}